\documentclass[12pt]{iopart}

\input{epsf} 

\usepackage{graphicx}

\newcommand {\bc} {\begin{center}}
\newcommand {\ec} {\end{center}}
\newcommand {\bd}{\begin{displaymath}}
\newcommand {\ed}{\end{displaymath}}
 \newcommand {\be} {\begin{equation}}
\newcommand {\bea} {\begin{eqnarray} \nonumber }
\newcommand {\ee} {\end{equation}}
\newcommand {\eea} {\end{eqnarray}}
 \newcommand {\eps} {\epsilon}

\newcommand {\Si} {\Sigma}
 \newcommand {\al} {\alpha}

\newcommand {\lan} {\langle}
\newcommand {\ran} {\rangle}

\newcommand {\cD}  {{\cal D}}

\def\eps{\epsilon}
\def\al{\alpha}

 \def\(({\left(}
 \def\)){\right)}
\def\[[{\left[}
\def\]]{\right]}
\def\bi{\bibitem}

\newcommand {\for} {\ \ \ \mbox{for}\ \ }

\def \form#1 {eq. (\ref{#1}) }
\def \parziale#1#2  {{\partial {#1} \over \partial {#2}}}

\begin{document}
\title{First Steps in Glass Theory}
\author{Marc M\'ezard}
\address{Laboratoire de Physique Th\'eorique de l'Ecole
Normale Sup\'{e}rieure  \footnote{Address after may 1st: Laboratoire de Physique 
Th\'eorique et Mod\`eles Statistiques, Universit\'e de Paris Sud, Bat. 100,
91405 Orsay, France}\\
24 rue
 Lhomond, F-75231 Paris Cedex 05, (France)\\
mezard@physique.ens.fr}

\begin{abstract}This paper is an introduction to some 
of the main present issues in
the theory of structural glasses. After recalling a 
few experimental facts, 
it gives a short account of the analogy between fragile 
glasses and the mean field
discontinuous spin glasses. The many valley picture is 
presented, 
and a brief account of recent attempts to  obtain  
quantitative
results from  first principle computations is summarised.
\end{abstract}

\section{Introduction}
When quenched fast enough so that it avoids the crystallisation
transition, almost any  liquid becomes a glass\cite{glass_revue}. This means that 
the density profile is not flat as in a liquid, it contains some
 peaks as in a crystal, but these peaks are not located on the nodes
 of a periodic or quasi periodic lattice. The understanding of such
amorphous 'solid' states has been recognised for a long time as a major
question in condensed matter physics. The sentence by Phil Anderson:
"... there are still fascinating questions of principle about glasses
and other amorphous phases..." \cite{PWA}, written nearly thirty years ago,
 was once again visionary in
that it foresaw the wonderful developments on glassy systems,
and particularly on spin glasses. The progress has been particularly difficult
in these area, and in particular as far as structural glasses are concerned.

\section{Mathematics}

The first question which comes to mind is whether  the glass is a new state 
of matter. It is not distinguished by any obvious symmetry (a not-obvious 
symmetry will be discussed later) from the liquid state, and one might think 
(as many
people do) that the density profile would actually become flat on time 
scales longer than the experimental ones: the glass would just be a 
liquid with 
a long relaxation time. 

 From the statistical physics point of view, one wants to start from 
a microscopic Hamiltonian. The simplest situation is that of  $N$ point-like
particles in a volume $V$, with a pair interaction potential
\be
H= \sum_{i<j} V_{ij}(r_i-r_j)
\ee
A simple case is that of  homogeneous systems where $V_{ij}$ is
independent of $i$ and $j$, and can be for instance either a hard sphere potential,
a `soft sphere' potential ($V_{ij}(r)=A/r^{12}$), or a Lennard-Jones
 potential ($V_{ij}(r)=A/r^{12}-B/r^6$). Also much studied
numerically\cite{kobrev}, because the crystallisation
is more easily avoided, are the binary mixtures where there are two types
of particles: each particle
$i$ has $\epsilon_i \in \{ \pm 1 \}$ and 
$V_{ij}(r)=V_{\epsilon_i \epsilon_j}(r)$, where $V_{++}$,
$V_{--}$, and $V_{+-}=V_{-+}$ are three potentials of the same type 
as before, but with different $A,B$  parameters corresponding to particles 
$+$ and $-$ having different
radii.

Does there exist, in any such case, an independent state of matter which 
is the glass state? Does it exist as a long-lived metastable state (like
the diamond phase of carbon)? Nobody knows the rigorous mathematical answer to
these questions. 
Actually much simpler related questions are unanswered (e.g. proving the
existence of a spin glass phase in a finite dimensional short range system
\cite{youngbook}),
or have taken many efforts to solve (e.g. proving Kepler's conjecture that
the densest three dimensional packing of hard spheres is the fcc/hcp lattice
\cite{kepler}).

\section{Experiments}

\begin{figure}
\includegraphics[width=0.40\textwidth,angle=270]{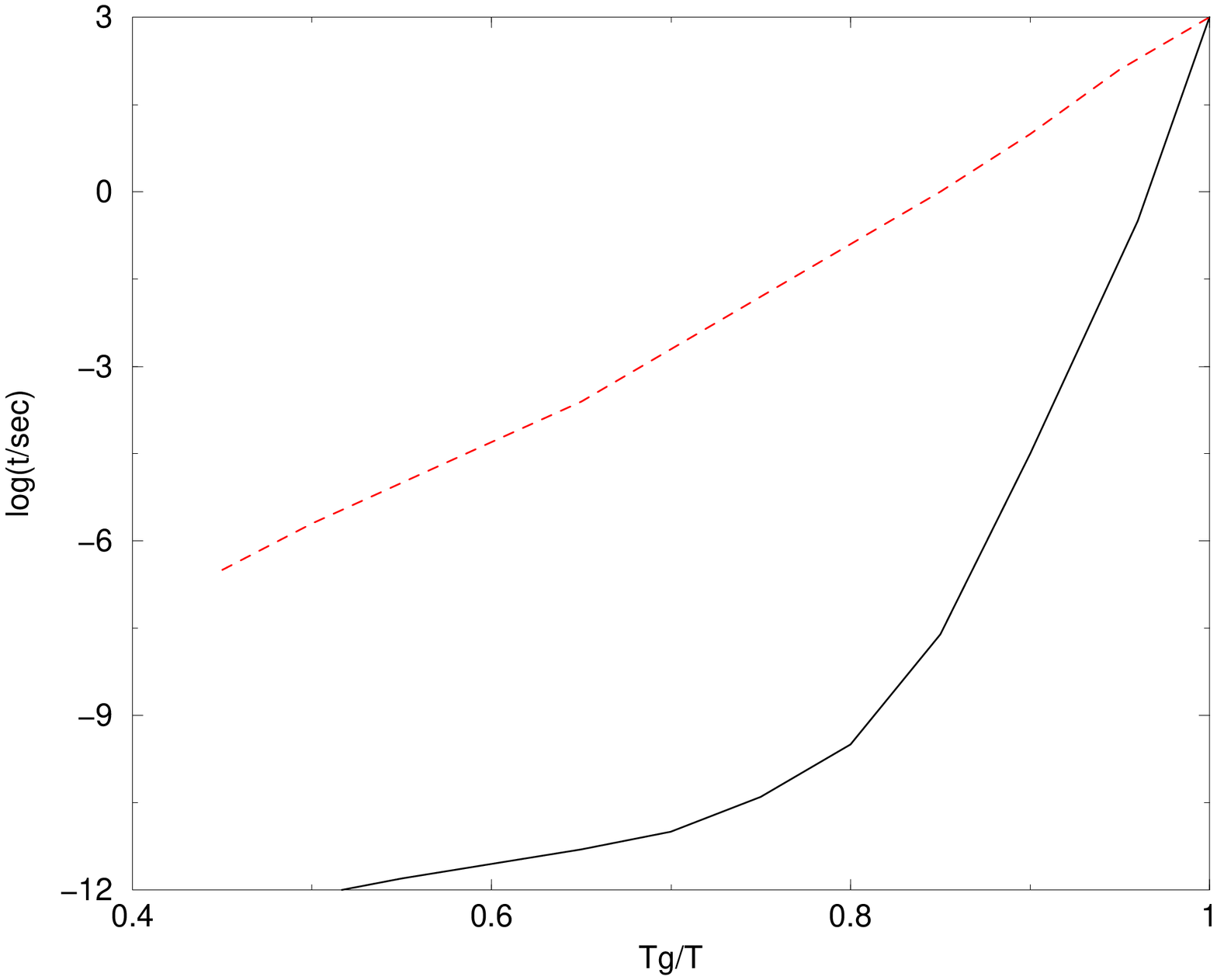}
\includegraphics[width=0.40\textwidth,angle=270]{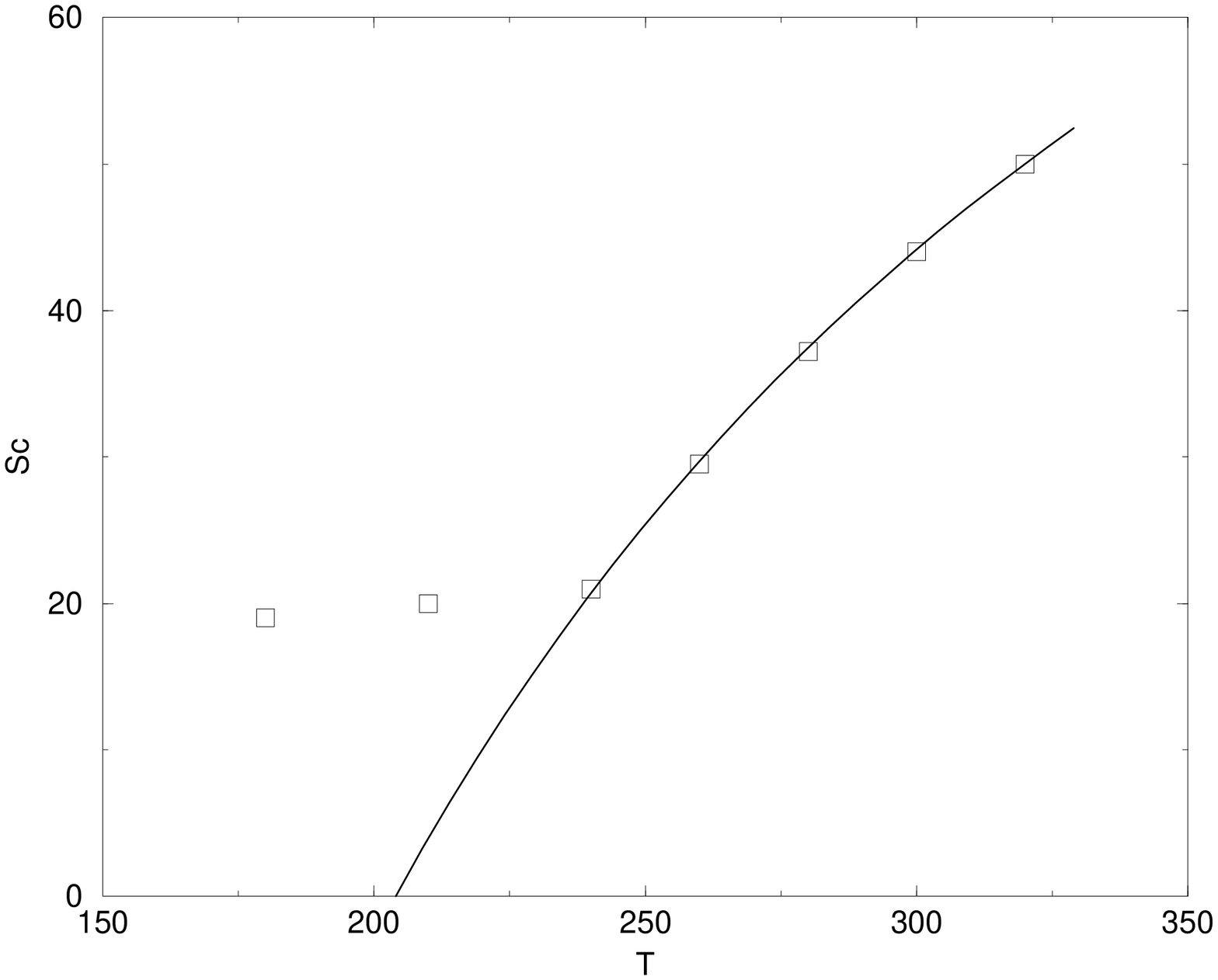}
\caption{The left-hand figure shows the behaviour of the logarithm ( in base 10) 
of the relaxation time in seconds,
versus $T_g/T$, for two extreme cases of glass formers. 
The dashed line is from GeO2, a strong glass former
which has an Arrhenius like behaviour;
The full line is the measurement from OTP, which is a fragile glass former
with a relaxation time diverging much faster than an Arrhenius law.
The right-hand figure shows the configurational entropy $S_c$ of  OTP (in $J K^{-1} mol^{-1}$)
versus temperature (in K). The configurational entropy, defined as the
difference between the entropy of the 
supercooled liquid and that of the crystal, is  measured through 
an integral of the specific heat difference. The squares are the experimental values.
The glass temperature is $T_g=246 K$. The full line is a fit to the equilibrated data,
of the type $S_c=S_\infty (1-T_K/T)$. The Kauzmann temperature is $T_K=204 K$, the
fusion temperature is $331 K$. The data are taken from \cite{RiAn}.}
\label{fig_Angell}
\end{figure}

Experimentally, the liquid falls out of equilibrium on experimental time scales, and 
becomes a `glass',
at a temperature $T_g$ called the glass temperature\cite{glass_revue}. This 
glass temperature is conventionnally defined as the one at 
which the relaxation time $\tau$ of the liquid, as obtained e.g. from
viscosity or from susceptibility measurements,
becomes of the order of $10^3$ seconds. Angell's plot of 
$\log\(( \tau /1 s \))$ versus $T_g/T$ allows to distinguish several types of behaviour (fig.
\ref{fig_Angell}).
So called strong glasses like $SiO_2$ have a typical Arrhenius behaviour 
with one well defined free energy barrier. 
On the other hand, some glasses, called fragile, show a dramatic increase of the relaxation time
when decreasing temperature which is much faster than Arrhenius:
the typical free energy barrier thus increases when $T$ decreases. This
implies a collective behaviour involving more and more particles.
An increase of the dynamical correlation, characteristic of the mobile
particles (rather than the more natural correlation of 
frozen particles), has been found in recent simulations \cite{benneman}.
A popular fit of the relaxation time versus temperature is the Vogel Fulcher one,
\be
\tau \sim \tau_0 \exp\(({A \over T-T_{VF}}\))
\ee
which would predict a phase transition at a temperature
$T_{VF}$ which is not accessible experimentally (while staying at equilibrium).
The more fragile the glass, the closer is $T_{VF}$ to $T_g$, while strong glass formers have a
$T_{VF}$ close to zero.

Another interesting experimental signature is that of the specific heat.
When one cools the liquid slowly, at a cooling rate $\Gamma=-dT/dt$, it 
freezes into a glass at a temperature
which decreases slightly when $\Gamma$ decreases. When this freezing occurs, the specific heat
jumps downward, from its value in the equilibrated supercooled liquid state to 
a glass value which is close to that of the crystal. From the specific heat, one can
compute the entropy.
The configurational entropy, defined experimentally as the difference $S_c=S_{liq}-S_{crystal}$,
behaves smoothly in the supercooled liquid phase, until the system 
becomes a glass (see fig.\ref{fig_Angell}). It was noted by Kauzmann long ago that, 
if extrapolated, $S_c(T)$ vanishes at
a finite temperature $T_K$. If cooled more slowly, the system
follows the smooth $S_c(T)$ curve down to slightly lower temperatures, but then freezes again.
 One can wonder what could happen at infinitely slow cooling. As a
negative $S_c$ does not make sense (except for pure hard spheres, where there is no energy),
 something must happen at temperatures above $T_K$. The curve $S_c(T) $ could flatten down smoothly,
or there might be a phase transition, which in the simplest
scenario would lead to $S_c(T)=0$ at $T<T_K$. This idea of an underlying "ideal"
phase transition, which
could be obtained only at infinitely slow cooling, receives some support from
the following observation: the two temperatures where the {\it extrapolated} experimental 
behaviour has a singularity,
$T_{VF}$ and $T_K$, turn out to be amazingly close to each other (see the  table below)\cite{RiAn}.
The first phenomenological attempts to explain this fact 
 originate in the work of Kauzmann \cite{kauzmann}, and 
developed among others 
by Adam, Gibbs and Di-Marzio \cite{AdGibbs}, which identifies the glass transition 
as a `bona fide'
thermodynamic transition blurred by some dynamical effects.

If there exists a true thermodynamic glass transition at $T=T_K=T_{VF}$,
it is a transition of a strange type.
It is of second order because the entropy and internal energy are 
continuous. 
 On the other hand the order 
parameter is
discontinuous at the transition, as in first order transitions: the
modulation of the microscopic density profile in the glass does not appear continuously from the
flat profile of the liquid. As soon as the system freezes, there is a finite jump
in this modulation (A more precide definition of the order parameter will be given below).

\vspace*{1cm}

{\bf{\centerline{Comparison of $T_K$ and $T_{VF}$ in various glass-formers (from \cite{RiAn})}}}

\noindent\begin{tabular}{p{5cm}p{3cm}p{3cm}p{3cm}}
Substance &$ T_K (K)$&$ T_{VF} (K)$&$ T_g (K)$\\[.2cm]
o-terphenyl&204.2&202.4&246\\[.2cm]
salol&175.2&&220\\[.2cm]
2-MTHF&69.3&69.6&91\\[.2cm]
n-propanol&72.2&70.2&97\\[.2cm]
3-bromopentane&82.5&82.9&108\\[.2cm]
\end{tabular}\\[.6cm]

\section{A mean field spin glass analogy}

A totally different class of systems where such a 1st-2nd order type transition was
found, and studied in great details, is a certain category of mean field spin glasses.
A few years after  the replica symmetry breaking (RSB) 
solution of the mean field theory of spin glasses \cite{MPV}, it was 
realized 
that there exists   another category of mean-field spin
glasses where the static phase transition exists and is due to an entropy crisis \cite{REM}. These
are now called discontinuous spin glasses because their phase transition, 
although
it is of second order in the Ehrenfest sense, has a discontinuous order
parameter \cite{GrossMez}.
 Another name often found in the literature is 
 `one step RSB' spin glasses, because of the
special pattern of symmetry breaking involved in their solution. 
These are spin glasses with infinite range interactions involving a coupling between triplets (or higher order
groups) of spins.
The simplest among them is the random energy model, which is the $p \to \infty$ limit
version of the p-spin models described by the Hamiltonian 
\be
H=-\sum_{i1<...<i_p} J_{i_1...i_p}
s_{i_1}...s_{i_p}
\ee
 where the $J$'s are (appropriately scaled) quenched random couplings,
and the spins can be either of Ising or spherical type \cite{GrossMez,KiThWo,crisanti}.

The analogy
between the phase transition of discontinuous spin glasses and the 
thermodynamic
glass transition was first noticed by Kirkpatrick, Thirumalai and Wolynes
in a series of inspired papers of the mid-eighties \cite{KiThWo}. While some of 
the basic ideas
of the present development were around at that time, there still missed a
few crucial ingredients. On one hand one needed to get more confidence 
that
 this analogy was not just fortuitous. 
The big obstacle was the existence (in spin glasses) versus
the absence (in structural glasses)  of quenched disorder. The 
discovery  of discontinuous spin glasses without any 
quenched disorder
\cite{nodis1,nodis2,nodis3}
provided an important new piece of information: contrarily to what had 
been
believed for long, quenched disorder is not necessary for the existence of
a spin glass phase (but frustration is).

\begin{figure}
\includegraphics[width=0.40\textwidth]{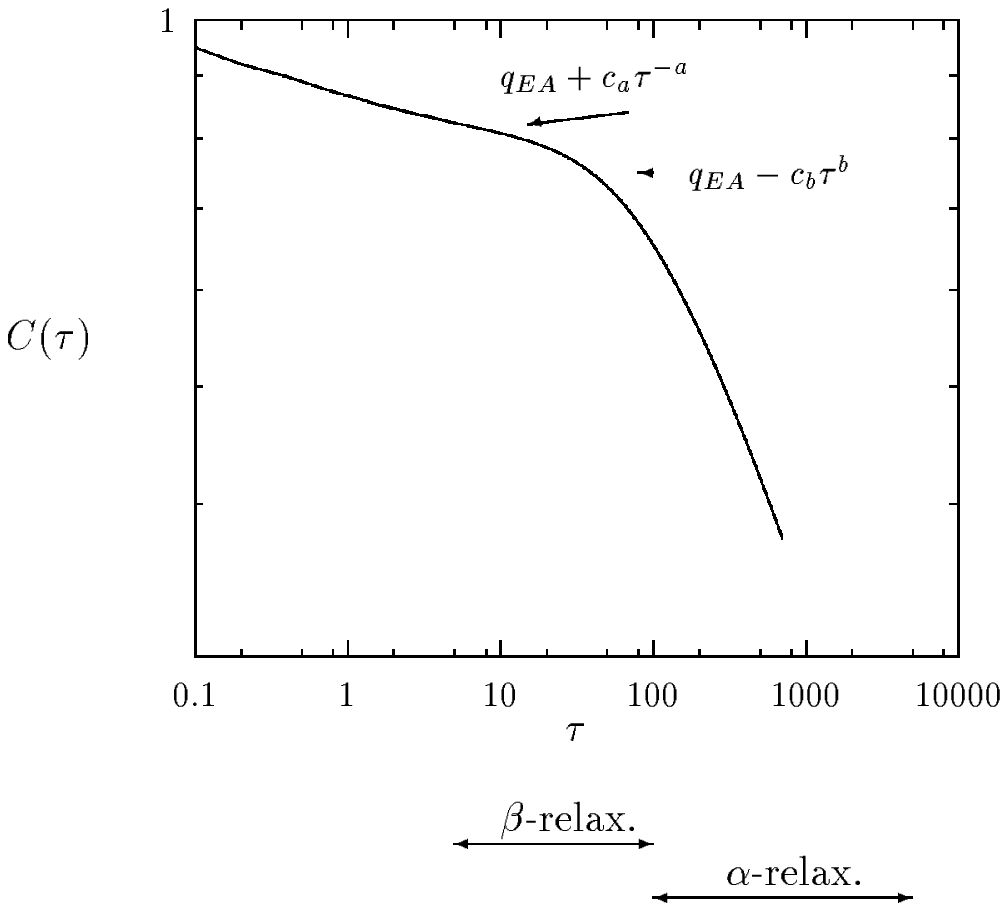}
\includegraphics[width=0.40\textwidth]{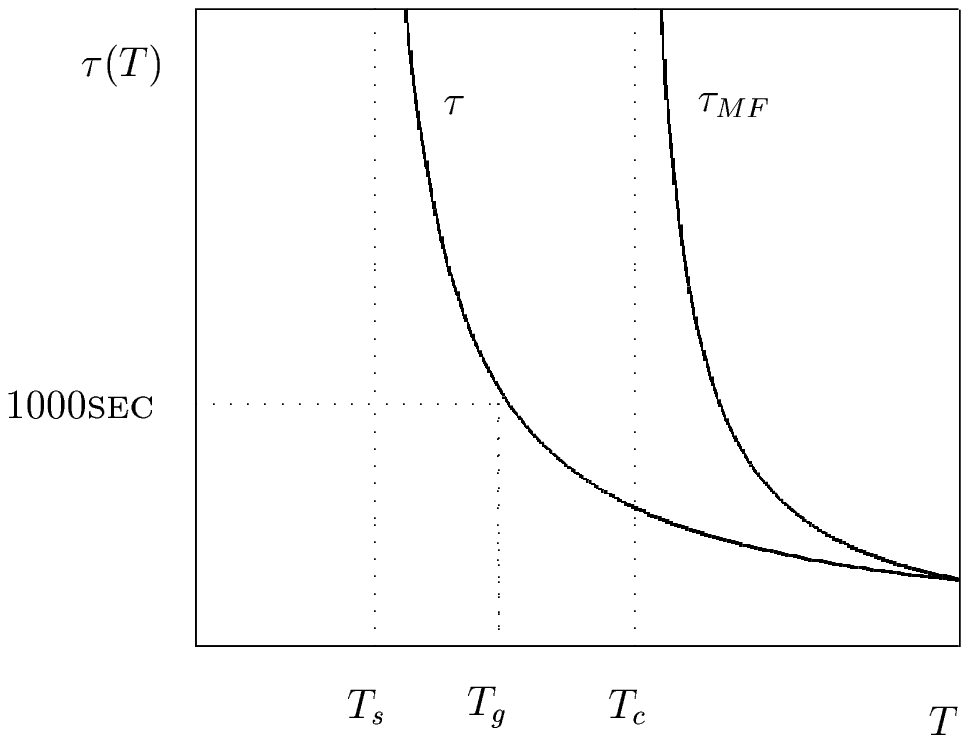}
\caption{The left-hand figure shows the schematic behaviour of the correlation function found in
mean field discontinuous spin glasses and observed in structural glasses.
The typical two-step relaxation consists of  a
fast $\beta$ relaxation leading to a plateau, followed by a
$\alpha$ relaxation from the
plateau, whose typical time scale  increases rapidly when $T$ decreases,
and diverges at $T=T_c$ which is equal to the
mode-coupling transition temperature. The right-hand figure shows the behaviour of the 
relaxation time versus temperature. The right-hand curve is the prediction
of mode-coupling theory without any activated processes: it is a mean field prediction,
which is exact for instance in
 the discontinuous mean-field spin glasses\cite{Bocukume}. The left-hand curve is the observed
relaxation time in a glass. The mode coupling theory 
provides a quantitative prediction for the increase
of the relaxation time when decreasing temperature, at high enough temperature (well above the
mode coupling transition $T_c$)\cite{gotze,MCexp}. 
The departure from the mean field prediction at lower temperatures is 
usually attributed to 'hopping' or 'activated' processes, in which the system is trapped
for a long time in some valleys, but can eventually jump out of it. The ideal
glass transition, which  takes place at $T_K$, 
cannot be observed directly since the system 
becomes out of equilibrium on laboratory time scales at the `glass temperature' $T_g$.
 }
\label{twostep}
\end{figure}

It is important to analyse critically  this analogy from 
the point of view of the dynamical behaviour. 
 In discontinuous mean field spin glasses there exist a dynamical transition temperature 
at a temperature $T_c$ which is larger than the equilibrium transition $T_K$. 
When T decreases and gets near to $T_c$,
the correlation function relaxes with a characteristic two step forms: a
fast $\beta$ relaxation leading to a plateau takes place on a characteristic time
which does not grow, while the $\alpha$ relaxation from the
plateau takes place on a time scale which diverges when $T \to T_c$ (see fig. \ref{twostep}). 
This dynamic
transition is exactly described by the schematic mode coupling equations. 

However the
existence of a dynamic relaxation at a temperature above the true thermodynamic one
is possible only in mean field, and the conjecture\cite{KiThWo}
 is that in a realistic system like a glass, 
the region between $T_K$ and $T_c$ will have instead a finite, but very rapidly 
increasing, relaxation time, as explained in fig. \ref{twostep}. A similar 
behaviour has been found in finite -size mean field models \cite{cririt}

 Another very interesting dynamical regime is the one where the
system is out of equilibrium ($T<T_g$). Then the system is no longer stationnary:
it ages. This is well known for instance from studies in polymeric glasses.
If one measures the response of your favorite plastic ruler to some stress,
it will behave differently depending on its age. Schematically, new
relaxation  processes
come into play on a time scale comparable to the age of the system: the older
the system, the longer the time needed for this "aging" relaxation to take place.
Recent years have seen  important developments
 on the out of equilibrium dynamics of the 
glassy phases \cite{BCKM},
initiated by the exact solution of the dynamics in a discontinuous spin 
glass
by Cugliandolo and Kurchan \cite{cuku}. It has become clear that, in realistic systems 
with short
range interactions, the pattern of replica symmetry breaking can be 
deduced
from the measurements of the violation of the fluctuation dissipation 
theorem \cite{fdr}.
These  measurements are difficult. However, numerical
simulations performed on different types of glass forming
systems have provided an independent and spectacular confirmation of their
`one step rsb' structure \cite{gpglass,bk1,bk2,leo} on the (short) time
scales which are accessible.
Experimental results have not yet settled the issue, but the first measurements of
effective temperatures in the fluctuation dissipation relation have been made  
recently \cite{fdr_exp}.

To summarize, the analogy between the phenomenology of fragile glass formers and 
discontinuous mean field spin glasses accounts for:
\begin{itemize}
\item The discontinuity of the order parameter
\item The continuity of the energy and the entropy
\item The jump in specific heat (and the sign of the jump)
\item Kauzmann's "entropy crisis"
\item The two steps relaxation of the dynamics and the succes of Mode Coupling 
Theory at relatively high temperatures.
\item
The aging phenomenon and the pattern of modification of the fluctuation dissipation relation
in the low temperature phase
\end{itemize}

\section{A lesson from mean field: many valleys}
\begin{figure}
\includegraphics[width=0.49\textwidth]{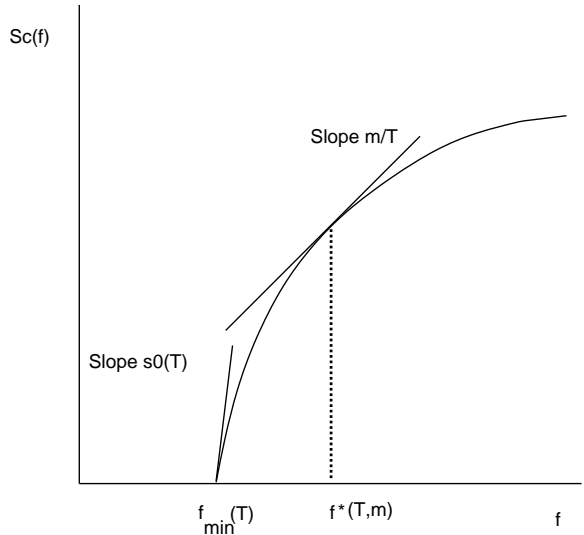}
\caption{
Qualitative shape of the complexity versus versus free energy
in  mean field discontinuous spin glasses. 
The whole curve depends on the temperature. The
saddle point which dominates the partition function,
for $m$ constrained replicas, is the point $f^*$ such
that the slope of the curve equals $m/T$ (for the usual unreplicated system, 
$m=1$).
If the temperature is small enough the saddle point sticks
to the minimum $f=f_{min}$ and the system is in its glass phase.
For $m=1$, this equilibrium phase transition happens at $T=T_K$.  }
\label{sigma_qualit}
\end{figure}

The successes of the above analogy suggest to have a closer 
look at the mean field models in order to understand, at least at the mean field 
level, what are the basic ingredients at work in the glass transition.
In mean field spin glasses, at temperatures $T_K<T<T_c$, the phase space breaks up into
ergodic components which are well separated, so-called free energy valleys or TAP
states \cite{TAP,crisomtap}. Each valley $\al$ has a free energy $F_\al$ and a free energy 
density 
$f_\al= F_\al/N$. The number of free energy minima with 
free energy density  $f$ is found to be exponentially large:
\be
{\cal N}(f,T,N) \approx \exp(N\Sigma(f,T)),\label{CON}
\ee
where the function $\Sigma$ is called the complexity. 
The total free energy of the system, $\Phi$, 
can be well approximated by:
\be
 e^{-\beta N \Phi} \simeq \sum_\al e^{-\beta N f_\al(T)} =
\int_{f_{min}}^{f_{max}} df \ \exp\((N[ \Sigma(f,T)-\beta f]\)) \ ,
\label{SUM}
\ee
where $\beta=1/T$.
The minima which dominate the
sum are those with a free energy density $f^*$
 which minimizes the quantity $\Phi(f)=f-T\Sigma(f,T)$.
At large enough temperatures the saddle point is at $f>f_{min}(T)$. When 
one
decreases $T$ the saddle point free energy decreases (see fig.\ref{sigma_qualit},
with $m=1$).
The Kauzman temperature $T_K$ is that below which the saddle point sticks
to the minimum: $f^*=f_{min}(T)$. It is a genuine phase transition \cite{REM,GrossMez,KiThWo}.
However because $T_c>T_K$, the phase space is actually separated into 
non ergodic components (valleys) at $T<T_c$ (actually there exist some non ergodic components
also above $T_c$, but they are not felt by the system
when starting from random initial conditions \cite{BaBuMez}). The total 
equilibrium free energy is analytic at $T_c$: in spite of the
ergodicity breaking, the system has the same free energy as that of the liquid,
as if transitions were allowed between valleys.

What remains of this mean field picture in finite dimensional glasses?
When one decreases the temperature, there is a well defined separation of 
time scales between the $\alpha$ and the $\beta$ relaxations, which suggest to
consider the dynamical evolution of system
in phase space  as a superposition of two processes: an intravalley (relatively fast)
relaxation, and an intervalley (slow, ang getting rapidly much slower when one cools the system)
hopping process. 

One popular way of making this statement more precise,
allowing  to study it numerically, is through
the use of inherent structures (IS) \cite{IS}. Given a configuration of
the system, characterized by its phase space position $x=\{ {\vec x_1},...,{\vec x_N}\}$, 
the corresponding inherent structure $s(x)$ is another point of phase space which is the
local minimum of the Hamiltonian which is reached from this configuration through a 
steepest descent dynamics. IS are easily identified numerically.
A given trajectory $x(t)$ of the glass through phase space maps onto the corresponding
trajectory $s(t)$ in the space of inherent structures. Looking at the
dynamical evolution in the space of IS \cite{schroder} makes the valley structure
slightly more apparent, since  one gets rid of the small thermal 
excitations around each valley minimum. Calling 
$\cD_{s}$ the set of those 
configurations which are mapped to the coherent structure  $s$,
a natural definition of the IS  entropy density, $\Si_{is}$, is
$
N \Si_{is}(T)= -\sum_{s} P(s)\ln(P(s))
$,
where the weight of the inherent structure $s$ is
\be
P(s)=Z(s)/\sum_b Z(b) \ \ \  ; \ \ \  Z(s)\equiv \int_{x \in \cD_{s} }dx \exp (-\beta H(x)) \ .
\label{za}
\ee

In a system with short range interactions, it is reasonable to expect that 
one may have two distinct IS which differ by a local rearrangement of
a finite number of atoms. It is then easy to show that the slope
of configurational
entropy versus free energy is infinite around $f_{min}$ \cite{Still_slope}, which does not
agree with the general scenario, except if the Kauzmann temperature vanishes.
This problem is due to te fact that IS are too simple objects, which cannot
 be identified with the free energy valleys. The difference is
very easily seen in spin systems \cite{BiMo}: IS are nothing but configurations
which are stable to one spin flip. Zero temperature free energy valleys, defined as TAP states,
are stable to the flip of  any arbitrarily large number $k$ of spins (but the
limit $N \to \infty$ must be taken before the limit $k \to \infty$). In continuous
systems, the generalization is clear: IS are local minima of the energy, so that any infinitesimal
move of the positions of all $N$ particles raises the energy. Let us
generalize the notion of a minimum as follows: define a k-th order local minimum as
a configuration of particles such that any infinitesimal move of all $N$ particles, 
together with a move of {\it arbitrary size} of $k$ particles, raises the energy. The limit
$k \to \infty$ gives the proper definition of a zero temperature free energy valley.
The proper definition at finite temperature
is slightly more involved  \cite{pot}. Let
us summarize it here briefly.
Given two configurations $x$ and $y$ we define their overlap as before as
$
q(x,y) = -1/N \sum_{i,k=1,N} w(x_{i}-y_{k}),
$
where $w(x)=-1 \for x$ small, $w(x)=0 \for x$ larger than the typical interatomic distance.
We add an extra term to the Hamiltonian:
we define
\bea
\exp (-N \beta F(y,\eps))=\int dx \exp( -H(x)+ \beta  \eps N q(x,y)),\\
F(\eps)=\lan  F(y,\eps) \ran ,
\eea
where $\lan f(y) \ran$ denotes the average value of $f$ over equilibrium configurations $y$
thermalized at temperature $\beta^{-1}$. Taking the thermodynamic limit
before the limit $\eps \to 0$ allows to identify the valley around any generic
equilibrium configuration $y$ \cite{pot,MP_Trieste}. 

In a nutshell, two configurations which differ by the (arbitrarily large) displacement
of a finite number of atoms are in the same thermodynamic valley. This definition of
the valleys also suffers from some difficulties: Nucleation
arguments then forbid the existence of a non-trivial
complexity versus free energy curve in a finite dimensional system. The
solution consists in noticing that
there exist many more metastable valleys, which have a finite but very long lifetime.
These can be identified by taking 
 the Legendre transform $W(q)$ of the free energy $F(\eps)$:
\be
W(q)=F(\eps)+\eps q \ \ \ ; \ \ \ q={-\partial F \over \partial \eps} \ .
\ee
Analytic computation in mean field models \cite{pot}, as well as in  glass forming liquids 
using the replicated HNC 
approximation \cite{card}, show that $W(q)$
 is minimal at $q=0$,
but has a secondary  minimum at a certain $q=q_{EA}$,
in the temperature range $T_K<T<T_c$. The behaviour around this second,
 metastable, minimum  
corresponds to phenomena that can be observed 
on time scales shorter than the lifetime of the metastable state.
The thermodynamic configurational entropy is the value of the potential
$W(q)$ at the secondary minimum with $q \ne 0$ \cite{pot}, and it
can be defined only if the minimum does exist (i.e. for $T<T_{c}$). 
Of course the secondary 
minimum for $T>T_{k}$ is always in the metastable region. 
 However if one would  start from a large value of $\eps$ 
and would  decrease $\eps$  to zero not too slowly,
the system would not escape from the metastable region and one 
obtains  a proper definition of the  thermodynamic configurational entropy in
this region $T>T_K$. In a similar way 
one could compute $q(\eps)$ in the region ($\eps > \eps_{c}$) where the high $q$ phase is 
thermodynamically stable
and extrapolate it to $\eps \to 0$.
The ambiguity in the definition of the thermodynamic configurational entropy 
at temperatures above $T_k$ becomes larger and 
larger when the temperature increases. It  cannot be defined for $T>T_{c}$.

\section{Beyond the analogy: first principles computation}
In recent years, it has become possible to go
beyond the simple analogy between structural glasses and mean field discontinuous spin glasses.
One can actually use the concepts and the techniques which
are suggested by this analogy in order to start a systematic first principles study of the
glass phase \cite{MePa1,MePa2,MP_Trieste}. 
So far we have focused onto the equilibrium study of the low 
temperature phase. 
One main reason is
that the direct study of  out of equilibrium dynamics is more difficult, 
and that one might be able
to make progress by a careful analysis of the landscape \cite{angelani}.
The strategy is to assume that there exists a phase transition, and that 
it is of the same type
as the one found in discontinuous mean field spin glasses. 
Within this framework, one tries to compute the properties of the glass phase. This
involves several 
 quantitites like the Kauzmann temperature, the radius of the cage which confine 
the particles in 
the glass phase, the configurational entropy etc... The validity of the scenario is checked
from the comparison of various predictions with numerical simulations of well 
equilibrated systems.

The first task is to define an order parameter. This is not trivial in an 
equilibrium theory where
we have no notion of time persistent correlations. The best way is to introduce two copies of
the system, with a weak interaction. The two sets of particles have
positions $x_i$ and $y_i$ respectively, the total Hamiltonian is
\be
E=\sum_{1 \le i \leq j \le N}( v(x_i-x_j) +v(y_i-y_j))+\eps \sum_{i,j} 
w(x_i-y_j)
\ee
where we have introduced a small attractive potential $w(r)$ between
the two systems. The precise shape of $w$ is irrelevant, insofar as we shall
be 
interested
in the limit $\eps \to 0$, but its range should be of order or smaller than
the 
typical 
interparticle 
distance. The order parameter is then the correlation function between the two 
systems:
\be
g_{xy}(r)=\lim_{\eps \to 0} \lim_{N \to \infty} 
{1 \over \rho N} \sum_{ij} <\delta(x_i-y_j-r)>
\ee
In the liquid phase this correlation function is identically equal to one,
while 
it
has a nontrivial structure in the glass phase, reminiscent of the pair 
correlation
of a dense liquid, but with an extra peak around $r \simeq 0$. Let us notice 
that we
expect a discontinuous jump of this order parameter at the transition, in spite
of its being second order in the thermodynamic sense. The existence of a non 
trivial order
parameter is associated with the spontaneous breaking of a symmetry: For 
$\eps=0$, with
periodic boundary conditions, the system is symmetric under a global
translation 
of
the $x$ particles with respect to the $y$ particles. This symmetry is 
spontaneously broken in 
the low
temperature phase, where the particles of each subsystem tend to sit in front of
each other.

Generalizing this approach to a system of $m$ coupled replicas,
sometimes named `clones' in
this context
(the order parameter used only $m=2$), provides a wonderful
method for studying  analytically the thermodynamics of the glass phase
\cite{remi,Me}.  In the glass phase, the 
attraction will force all $m$ systems
 to fall into the same glass state, so that
the  partition function is:
\be
Z_{m} = \sum_\al e^{-\beta Nm f_\al(T)}= \int_{f_{min}}^{f_{max}} df
 \ \exp\((N [ \Sigma(f,T)-m \beta f]\))
\label{zm}
\ee
In the limit where $m \to 1$ the corresponding partition function 
$Z_m$ is dominated by the correct saddle point $f^*$ for $T>T_K$. 
The interesting regime is when the temperature is  $T<T_K$, 
and the number $m$ is allowed to become smaller than one. The saddle 
point $f^*(m,T)$ in the expression (\ref{zm}) is the solution
of $\partial \Sigma(f,T) / \partial f=m/T$. Because of the
convexity of $\Sigma$ as function of $f$, the saddle point is
at $f>f_{min}(T)$ when $m$ is small enough, and it   sticks at 
$f^*=f_{min}(T)$
when $m$ becomes larger than a certain value $m=m^{*}(T)$,
a value which is smaller than one when $T<T_K$ (see fig. \ref{sigma_qualit}). 
The free energy
in the glass phase, $F(m=1,T)$, is equal to $ F(m^*(T),T)$. As the free 
energy
is continuous along the transition line $m=m^*(T)$, one can compute 
$F(m^*(T),T)$ from the region $m \le m^*(T)$, which is a region where the
replicated system is in the liquid phase. This is the clue to
the explicit computation of the free energy in the glass phase. 
It may sound a bit strange because one is tempted to think of $m$ as an 
integer
number. However the computation is much clearer if one sees $m$ as 
a real parameter in (\ref{zm}). As one considers low temperatures $T<T_K$ 
the
$m$ coupled replicas fall into the same glass state and thus they build
some molecules of $m$ atoms, each molecule being built from one atom of 
each 
'colour'. Now  the interaction strength of one such molecule with another 
one
is basically  rescaled by a factor $m$ (this 
statement becomes  exact in the limit of zero temperature
where the molecules become point like). If $m$ is small enough this 
interaction is small
 and the system of molecules is liquid. When $m$ increases, the molecular 
fluid
freezes into a glass state at the value $m=m^*(T)$.
So our method requires to estimate the 
 replicated free energy, 
$
F(m,T)=-{\log(Z_m) /( \beta m N )}
$,
 in a molecular
liquid phase, where the molecules consist of $m$ atoms and
$m$ is smaller than one. For $T<T_K$, $F(m,T)$ is maximum at
the value of $m=m^{*}$ smaller than one,
while for $T>T_K$ the maximum is reached at a  value $m^*$ is larger than one.
  The knowledge of $F_m$ as a 
function
of $m$ allows to reconstruct the configurational entropy
function $Sc(f)$ at a given temperature $T$
through a Legendre transform, using the parametric representation (easily
deduced from a saddle point 
evaluation of (\ref{zm})) \cite{remi}:
\be
f={\partial \[[m F(m,T)\]] \over \partial m} \ \ \ ; \ \  \Sigma(f)={m^2 \over T}
{\partial F(m,T)\over \partial m} \ .
\label{legend}
\ee

 The Kauzmann temperature ('ideal 
glass
temperature') is the one such that $m^*(T_K)=1$. For $T<T_K$ the equilibrium
configurational entropy vanishes. Above $T_K$ one obtains the equilibrium
configurational entropy $\Sigma(T)$ by solving (\ref{legend}) at $m=1$. 

This gives the main idea which allows to compute the free energy in the
glass phase, at a temperature
$T<T_K$, from first principles: it is equal to the free energy of a molecular liquid
at the same temperature, where each molecule is built of $m$ atoms, and
an appropriate analytic continuation to $m=m^*(T)<1$ has been taken. The whole problem 
is reduced to a computation in a liquid. This is not trivial, and requires to develop
some specific approximations. I shall not elaborate on that here, but 
refer the reader to the original papers \cite{MePa2,sferesoft,LJ,LJ2}.
 The basic idea of the approximation  is that the size of the molecules
is directly related to the thermal wandering of an atom in its cage.
Therefore at low temperatures one can use some small cage approximation.
it is natural to write the partition function in terms
of the center of mass and relative
coordinates $\{ r_i, u_i^a \}$, with $x_i^a=r_i+u_i^a$ and $\sum_a u_i^a=0$, 
and to
expand the interaction in powers of the relative displacements $u$.
Keeping only the term quadratic in $u$ (harmonic vibrations of the molecules), and integrating
over
these vibration modes, one gets the "harmonic resummation"
approximation where the partition function is given by:
\be
Z_m= Z_m^0  \int dr
\exp\((-\beta m H(r) -{m-1 \over 2} Tr \log  M \)) 
\label{Zharmo}
\ee
where $ Z_m^0 ={m^{Nd/2} \sqrt{2 \pi T}^{N d (m-1)} / N!}$, and
the matrix $M$, of dimension $dN \times dN$, is given by:
\be
M_{(i \mu) (j \nu)}= {\partial^2 H(r) \over \partial r_i^\mu \partial r_j ^\nu}
= \delta_{ij} \sum_k v_{\mu\nu}(r_i-r_k)-  
v_{\mu\nu}(r_i-r_j)
\ee
and $v_{\mu\nu}(r) =\partial^2 v /\partial r_\mu \partial r_\nu$ 
(the indices $\mu$ and $\nu$ denote space directions).
Now we are back to a real problem of liquid theory, since we have only $d$ degrees of freedom
per molecule (the center of mass coordinates), and the number of clones, $m$, appears
as a parameter in (\ref{Zharmo}).

Once one has derived an expression for the replicated free energy, one can
deduce from it the whole thermodynamics, as described above (Notice
that the `technical' approximation  of neglecting the exchange of atoms between 
different molecules, as well as using a harmonic model, means that one really studies the
IS in this computation, rather than the real free energy valleys).
In all three cases, one finds an estimate of the Kauzmann temperature which is in 
reasonable agreement with simulations, with a jump in specific
heat, from a liquid value at $T>T_K$ to the Dulong-Petit value
$C=3/2$ (we have included only positional
degrees of freedom) below $T_K$.
This is similar to the experimental result, where the glass
specific heat jumps down to the crystal value when one decreases the temperature
(Our approximations so far are similar to the Einstein approximation of
independent vibrations of atoms, in which case the contribution
of positional degrees of freedom to the crystal specific heat is $C=3/2$).
The parameter $m^*(T)$ and the cages sizes
are nearly linear with temperature in the whole glass phase. 
This means, in particular, that the effective temperature $T/ m$ is always 
close to $T_K$, so in our theoretical computation we need
only to evaluate the expectation values of observables in the liquid phase, 
at temperatures where the HNC approximation for the liquid still works quite well. 

A more detailed numerical check of these analytical predictions involves
the measurement of the  complexity,
\be
\Sigma_{t}=S(T)-S_{valley}(T)
\label{Stdef}
\ee
 The liquid entropy is estimated 
by a thermodynamic integration of the specific heat from the very dilute 
(ideal gas) limit. It turns out that in the deeply supercooled region
the temperature dependence of the liquid entropy is well fitted by the
law predicted in \cite{taraz}: $S_{liq}(T)=a T^{-2/5}+b$, which presumably allows
for a good extrapolation at temperatures $T$ which cannot be simulated.
As for the 'valley' entropy, it can be estimated as that of an harmonic
solid. One needs however the vibration frequencies of the solid. These have been 
approximated by several methods, most of which are based on some
evaluation of the Instantaneous Normal Modes (INM) \cite{INM} in
the liquid phase,
and the
assumption that the spectrum of frequencies does not depend much on
temperature below $T_K$. Starting from a typical configuration
of the liquid, one 
can look at the INM  around it. In general there exist some negative eigenvalues
(the liquid is not a local minimum of the energy) which one must take care of. 
Several methods have been tried: either one
keeps only the positive eigenvalues, or one considers the absolute values of the eigenvalues
\cite{sferesoft,LJ,LJ2}.
Alternatively one can also consider the INM around the nearest inherent structure
which has by definition a positive spectrum \cite{sferesoft,LJ,LJ2,SKT}.
This procedure really measures the configurational entropy rather than the
thermodynamic complexity. The
computation of the thermodynamic  complexity, using its definition as
a system  coupled to a reference
thermalized configuration, has also been computed in \cite{sferesoft}
and turns out to be not very different from the configurational
entropy, on the time and temperature scales which have been studied so far
(they must differ on infinitely long time scales, as we discussed in the previous section). 

\begin{figure}
\includegraphics[width=0.35\textwidth,angle=270]{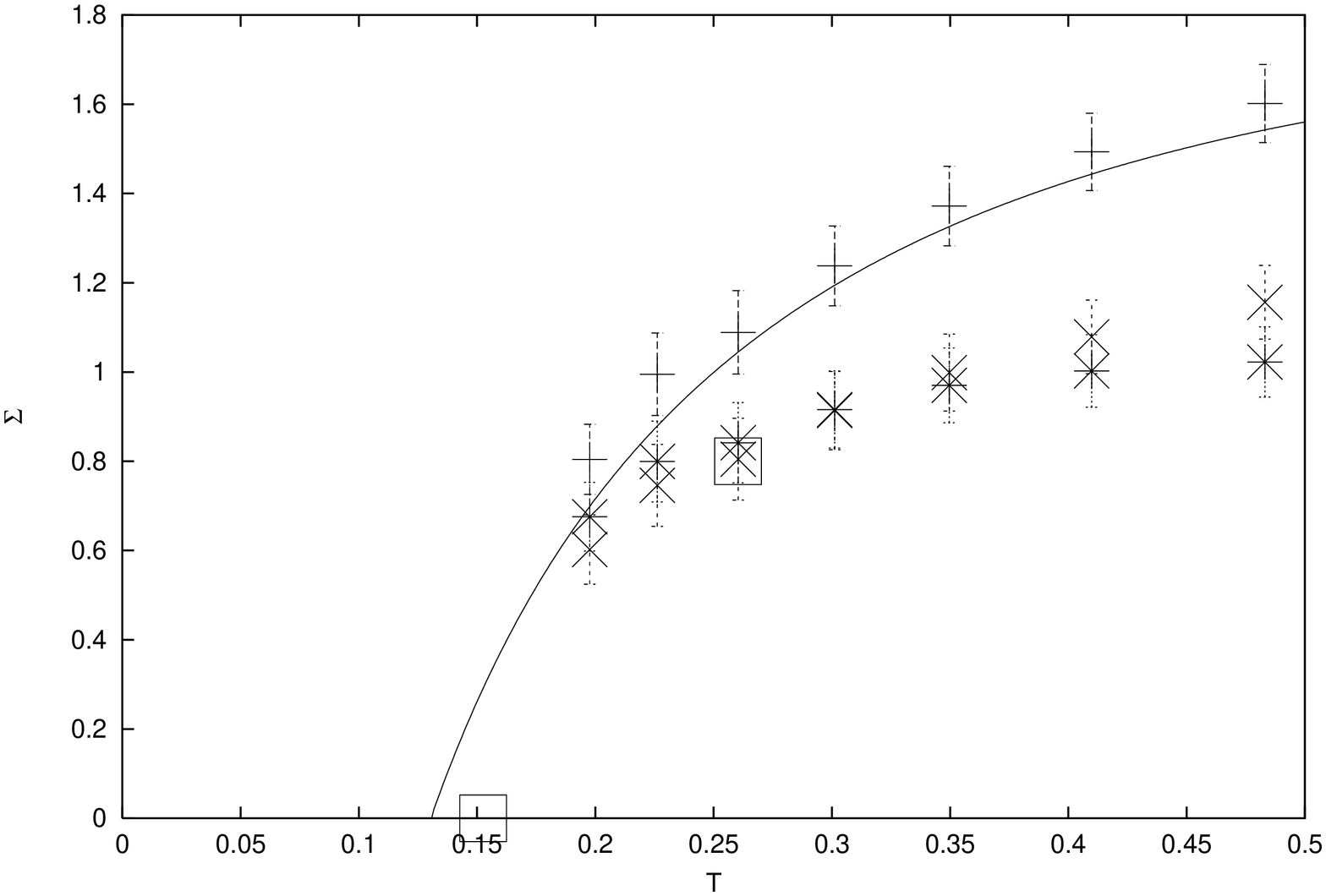}
\includegraphics[width=0.40\textwidth,angle=270]{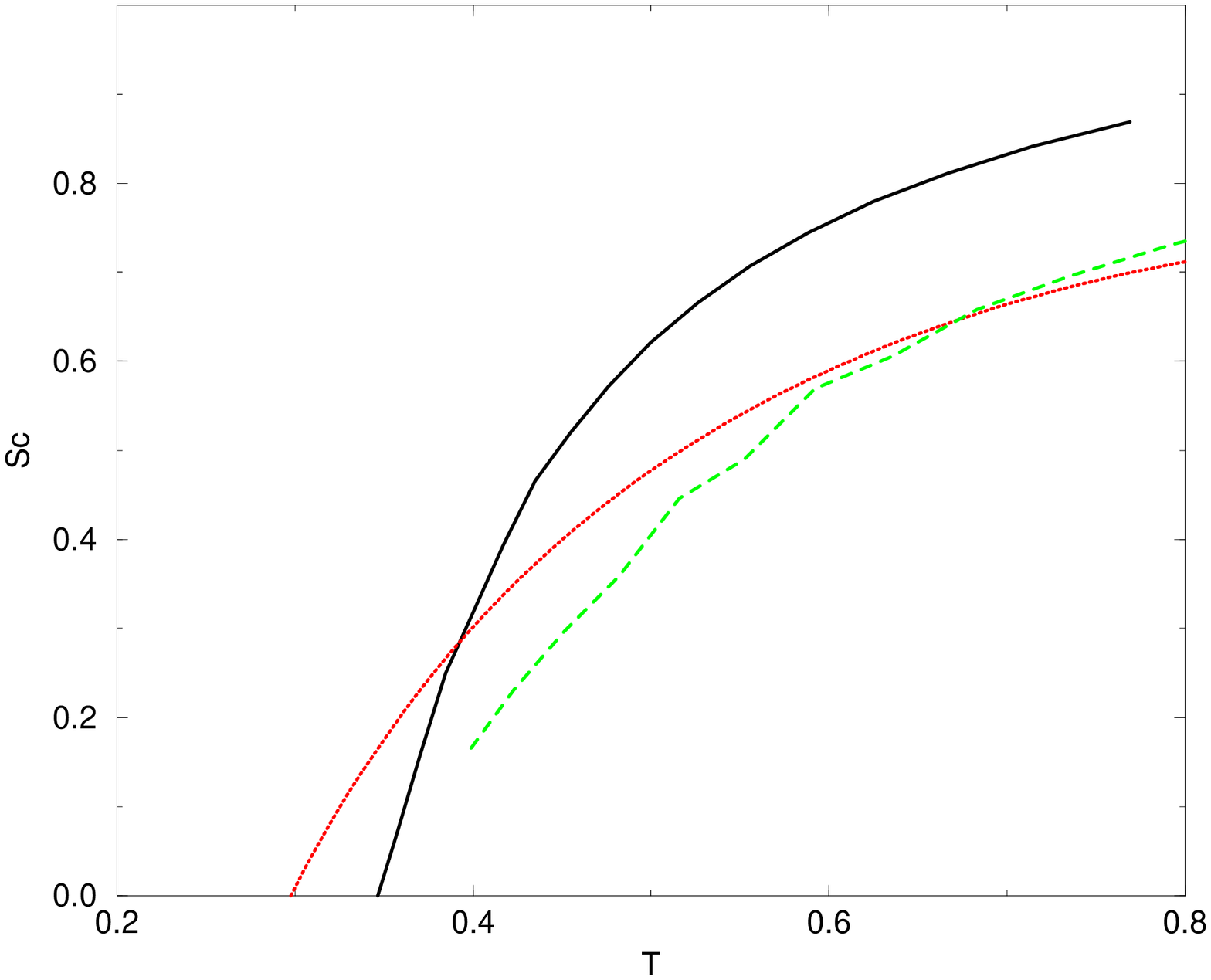}
\caption{The configurational entropy versus temperature in 
binary mixtures of soft-spheres and of   Lennard-Jones particles.
The soft sphere result (left curve), from \cite{sferesoft}, compares the analytical prediction
obtained within the harmonic resummation scheme (full line), to simulation
estimates of $S_{liq}-S_{valley}$, where the valley entropy is 
that of a harmonic solid with INM eigenvalues projected onto positive
eigenvalues (+), taken in absolute values ($\times$), or taken around
the nearest inherent structure ($\ast$). The  squares
correspond to
the numerical estimate of the
thermodynamic configurational entropy obtained by studying the system coupled  
to a reference configuration (see text, and \cite{sferesoft} for details). 
The Lennard-Jones result (right curve), shows as a 
full (black) curve the theoretical prediction obtained from the
cloned molecular liquid approach\cite{LJ,LJ2}. The dotted (green) curve is the result from
the simulations of \cite{LJ,LJ2} and the dashed (red) curve is the result
from the simulations of \cite{SKT}. Both simulations use the $S_{liq}-S_{valley}$
estimate where the harmonic solid vibration modes are approximated by the ones
of the nearest inherent structure.}
\label{figSc}
\end{figure}

The results for the configurational entropy as a function of temperature are
shown in fig.\ref{figSc}, for binary mixtures of soft spheres and of Lennard-Jones particles. 
The agreement with the analytical result obtained from the replicated fluid system
is rather satisfactory, considering the various approximations
involved both in the analytical estimate and in the numerical ones.

\section{Conclusion}
Our knowledge on first principle computations of glasses is still rather
primitive. Basically we have obtained, for the glass, the equivalent of the
Einstein approximation for the crystal. Even within this simple scheme,
doing the actual computation for the glass turns out to be rather involved.
What is most needed next is: on the analytical side,
some better approximations of the molecular liquid state, allowing to go beyond the
small cage expansions, and some reliable estimation of time scales in the
regime $T_K<T<T_c$;  on the numerical side, some
precise  results in the glass phase at equilibrium \cite{krauth}; 
on the experimental side, 
some more  measurements of the fluctuation dissipation ratio
in the  out of equilibrium dynamics.
No doubt:
"... there are still fascinating questions of principle about glasses
and other amorphous phases..." \cite{PWA}.

\section{Acknowledgments}
\label{acknowledgements}
It is a great pleasure to thank Giorgio Parisi for the 
 collaboration which led to the works described here,
as well as G.Biroli and R. Monasson for useful discussions.

\end{document}